\documentclass[10pt,preprint2]{aastex}

\shorttitle{}
\shortauthors{Mullally et al.}

\newcommand{\mj}{\,$\mathrm{M_J}$}
\newcommand{\au}{\,AU}
\newcommand{\um}{\,$\mu$m}

\newcommand{\msolar}{\,M$_{\odot}$}

\newcommand{\logg}{$\log{g}$}

\begin{document}

\title{Spitzer Planet Limits around the Pulsating White Dwarf GD66}
\author{Fergal Mullally\altaffilmark{1}, William T. Reach\altaffilmark{2}, Steven Degennaro\altaffilmark{3}, Adam Burrows\altaffilmark{1}}

\altaffiltext{1}{Department of Astrophysical Sciences, Princeton University, Princeton, NJ 08544; fergal@astro.princeton.edu}
\altaffiltext{2}{Spitzer Science Center, MS 220-6, California Institute of Technology, Pasadena, CA 91125}
\altaffiltext{3}{Department of Astronomy, University of Texas at Austin, Austin 
TX 78712}

\begin{abstract}
We present infrared observations in search of a planet around the white dwarf, GD66. Time-series photometry of GD66 shows a variation in the arrival time of stellar pulsations consistent with the presence of a planet with mass $\geq$2.4\mj. Any such planet is too close to the star to be resolved, but the planet's light can be directly detected as an excess flux at 4.5\um.  We observed GD66 with the two shorter wavelength channels of IRAC on Spitzer but did not find strong evidence of a companion, placing an upper limit of 5--7\mj\ on the mass of the companion, assuming an age of 1.2--1.7\,Gyr.


\end{abstract}


\section{Introduction}
The advantage of searching for sub-stellar companions around white dwarf stars (WDs) has long been recognized. The improved contrast between a WD and a companion was first exploited by \citet{Probst83} and the first L dwarf around a WD (or indeed any star) was discovered around GD165 by \citet{Becklin88}. In recent years, improved instrumentation has enabled searches for lower mass objects, to the point where direct detection searches for planetary mass objects are possible from the ground \citep[e.g.][Hogan et al. 2008, in press]{Burleigh02, Debes05cc2}. Searches utilizing the {\it Spitzer} Space Telescope \citep{Werner04} can obtain mass limits as low as $\sim$10\mj\, \citep[][]{Farihi08planet, Mullally07thesis, Debes07cc4}. 

WDs are interesting targets for more reasons than just their luminosity. By looking at WDs we can sample a range of progenitor masses from about 1-8\msolar, including higher mass stars not amenable to radial velocity surveys. Currently, the best main-sequence targets for direct detection of planetary companions are either transiting systems \citep[e.g.]{Charbonneau05, Deming06} or very young stars \citep{Marois08,Kalas08}. Detection of a planet around a WD would open up a older, more sedate region of parameter space where the predictions of planet models could be tested against observation for ages ranging from hundreds of millions to billions of years. WD planets are key to testing our models of planetary evolution. 

WDs also lend insight into the ultimate fate of planetary systems. \citet{Livio84} first estimated the conditions necessary for a gas giant planet to survive the red giant phase of its parent star. \citet{Sackmann93} concluded that the Earth would survive the red giant phase of the Sun, but more recent calculations by \citet{Schroder08} concluded the opposite. These calculations depend on stellar mass loss rates that are weakly constrained by observation due to the rapid evolution of evolved stars. Determining the distribution of planets around WDs will provide a key test for these models. The winds from evolved stars pollute the interstellar medium with the material that creates the next generation of terrestrial planets, so surveys for planets around dead stars will inform our understanding of planets around stars not yet born.

Early empirical success comes from the detection of a planet around a sub-dwarf star using pulsation timings \citep{Silvotti07}, planets in an eclipsing sub-dwarf system \citep{Lee08} and the detection of silicate rich debris disks around WDs \citep[e.g.][]{Reach05b, Farihi07gd362}. Analysis of the composition of material raining down onto the surface of a WD from such a disk by \citet{Zuckerman07} concluded that the pattern of elemental abundances is consistent with a species of asteroid. Asteroids are markers of planetary systems, and so the frequency of debris disk WDs may eventually provide an independent measure of the fraction of stars that bear (or bore) planetary systems. 

\citet[][hereafter M08]{Mullally08} published the results of a survey that sought to detect evidence of planets as variations in the arrival time of pulsations from variable WDs (in a manner analogous to sub-dwarf and pulsar planets). For one object, GD66, they discovered a variation consistent with a 2.1\mj\ planet in a 4.5\,yr orbit at a distance of about 50\,pc. Unfortunately, the span of their data was not long enough to cover an entire orbit and their detection remains provisional. In this paper we report on follow up observations with {\it Spitzer} in an attempt to confirm this object by direct detection in the infrared. Although the candidate system is too distant to resolve into individual components, the low luminosity of the star means that the flux from the planet at 4.5\um\ rivals that of the WD. 
For wavelengths less than 10\um  atmosphere models of planets are brightest at 4--5\um, in a gap between absorbption bands of methane and water \citep[e.g][]{Burrows03}. The flux from this bump increases with mass, and decreases with age, and provides a method of both detecting a planet and determining its mass.
The goal of these observations is to measure an excess flux at 4.5\um\ over the bare photosphere of GD66 due to the light from the planet. We achieve this by measuring the flux ratio (i.e. the color) between channels 2 and 1 on IRAC \citep[4.5 and 3.6\um\ respectively,][]{Fazio04} and comparing this ratio to a sample of other similar stars. This approach is capable of detecting companions independent of distance to the star, or the assumptions of atmosphere models.

These observations push the IRAC instrument in a new direction. While studies of transiting planets with IRAC have been able to detect changes in flux in a single band of order a few parts in $10^{-3}$, our goal was to measure an absolute flux excess relative to other wavelengths. With this approach, we established a sensitivity limit of $\sim$0.5\% with the two IRAC bands available during the warm mission.

\section{Observations}
We observed \object[WD0517+307]{GD66} with all four IRAC channels on 17th October 2007 (AORs 22855424 and 22856448) as part of program 40255. We recorded sixteen 100 second exposures  simultaneously in channels 1 and 3, using a spiral dither pattern. We repeated the procedure for channels 2 and 4, recording 4 sets of sixteen images for a total exposure time of 6,400 seconds. The raw frames were calibrated using version S16.10.0 of the IRAC pipeline to create bcd frames. 

We also observed 3 WDs with similar temperatures to calibrate our observations. 
\object[WD1425-811]{L19-2} (AORs 22856192 and 22857216), \object[WD0133-116]{ZZ Ceti} (AORs 22855936 and 22856960) and \object[WD1350+656]{G238-53} (AORs 22855680 and 22856704) are all DAVs, and \citet{Bergeron04} reports temperatures within 120\,K of GD66 based on optical spectroscopy. This agreement is better than the quoted uncertainty in the measurements, and so we can consider our sample to have a uniform temperature. As an independent, albeit cruder approach, we also insisted on broadly similar pulsation properties for each star, as \citet{Mukadam06} showed that pulsation properties are correlated with temperature. We observed each of these stars in a similar pattern as GD66, scaling our total exposure time with magnitude. 

We extracted photometry from these frames using the astrolib package in IDL. We first convert the recorded flux into units of electrons. Pixels with bits set in any of the pmask, imask or rmask images were marked as bad and their values set to the median value of pixels in a box 21 pixel on a side centered on the bad pixel in question. We ignored bit 13 of the imask, indicating crosstalk, which was set in a large number of pixels that did not seem to be affected by any noticeable problem. \citet{Graham90dust} \& \citet{Reach08} see evidence of pulsation at infrared wavelengths in G29-38, a WD of similar temperature but also possessing a close-in debris disk \citep{Reach05b}. However, we find no evidence of pulsation or any systematic drift in the measured flux in any channel for any star in our sample.

We extract the flux using a sky annulus of 10-20 pixels and series of apertures ranging from 2 to 7 pixels radius. We choose the aperture that gives the greatest signal to noise (2 pixels), adding in quadrature a contribution to the error term due to the read noise, as listed in the header of the bcd frame. We then apply corrections to the observed flux as suggested by the Data Handbook: array location dependence, aperture correction, color correction and a pixel phase correction for Channel 1. 

\section{Results}
\begin{figure}
    \begin{center}
	\includegraphics[angle=270, scale=.5]{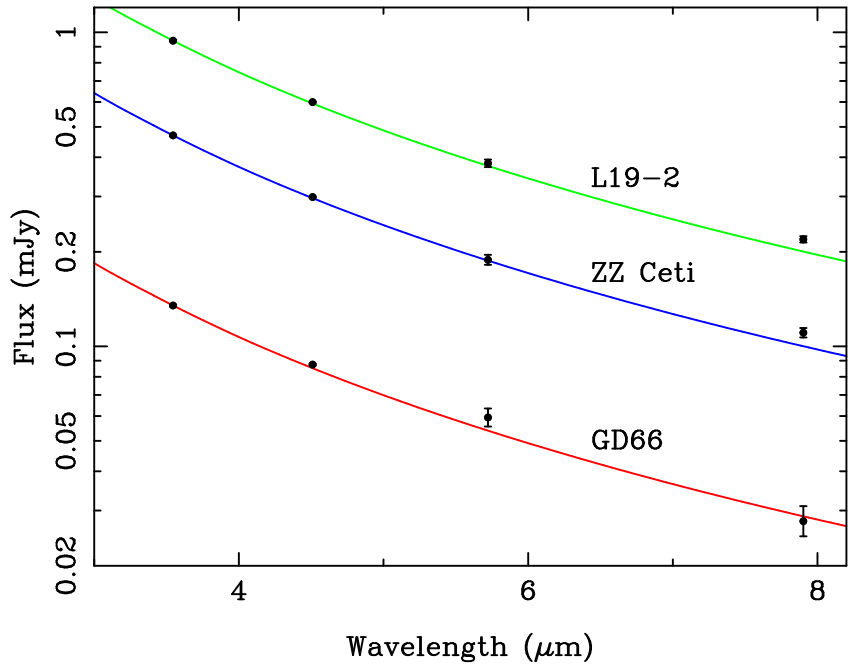} 
	\caption{Fluxes of three WDs in all 4 IRAC bands. The solid line connecting the points is a 12,000\,K DA white dwarf atmosphere model. The model is normalised to agree with the flux at 3.6\um and is shown to guide to the eye.\label{GD66sed}}
    \end{center}
\end{figure}

\begin{figure}
    \begin{center}
	\includegraphics[angle=270, scale=.5]{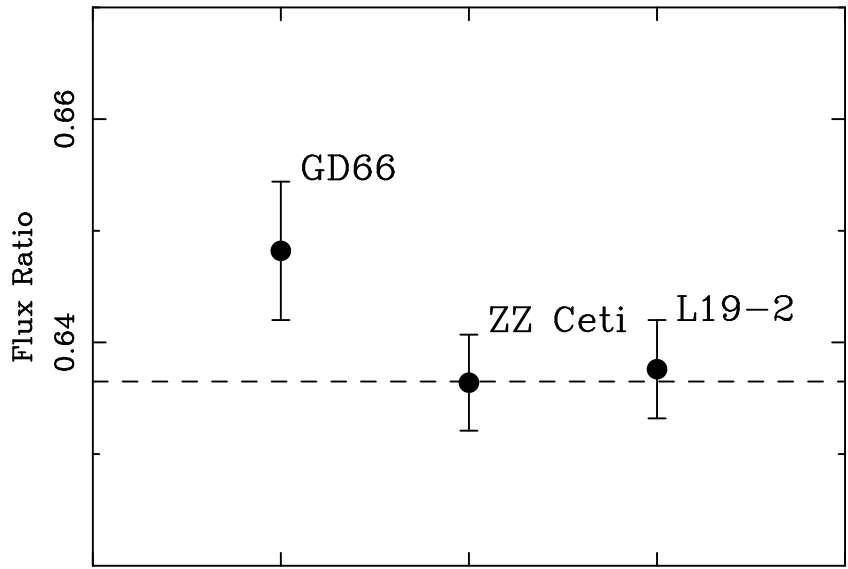}
	\caption{Flux ratio between channel 2 and channel 1 for three WDs. The dash line is the weighted average flux of the two comparison WDs.\label{xsflux}}
    \end{center}
\end{figure}

In Figure~\ref{GD66sed} we show spectral energy distributions (SEDs) for GD66, L19-2 and ZZ~Ceti. We show a table of observed fluxes for all observed stars in Table~\ref{allfluxes} and the flux ratio between channels 2 and 1, $r$, for each star in Figure~\ref{xsflux} and Table~\ref{fluxratio}. 

GD66 and G238-53 are noticeably redder (higher value of $r$) than the other two stars. However, careful analysis of the images shows that this flux excess is most likely an artefact for G238-53. The diffraction spike from a nearby bright star passes very close to the position of G238-53 on the chip. Unlike the other three stars, the observations for the two channels for this star were not take sequentially, but separated by $\sim$ 1 week. The slightly different viewing angle between the two visits means that the diffraction spike is closer to the star in the epoch of channels 2 and 4 (which are observed simultaneously). Although some of these pixels are masked, we consider the measured fluxes to be untrustworthy. The flux ratio is significantly higher (0.675 versus 0.657) for a 3 pixel aperture, a further indication that flux is contaminated. We therefore ignore G238-53 in our analysis.

\subsection{Error Analysis}
We now determine whether this excess in $r$ for GD66 is statistically significant. The error budget of an IRAC observation is divided to photometric, systematic and calibration components.  The IRAC data handbook gives a value of 3\%
for the calibration uncertainty while \citet{Farihi08dust} concludes, based in part on a literature survey that 5\%
is more appropriate.
However, because we are measuring a flux ratio, we need not concern ourselves with the absolute calibration error in converting from electrons to mJy necessary to compare IRAC fluxes to other photometric systems. We do, however, include a conservative uncertainty term of 5\%
 in Table~\ref{allfluxes} to facilitate comparison with other stars.

The photometric uncertainty in channel 1 can be computed from the rms scatter of individual bcd frames. Adding the fractional uncertainties in quadrature gives the fractional uncertainty in the flux ratio.
Our observations were designed to achieve $<$1\% photometric accuracy, and this goal was comfortably exceeded, as shown in Table~\ref{fluxratio}.  

\begin{figure}
    \begin{center}
	\includegraphics[angle=270, scale=.45]{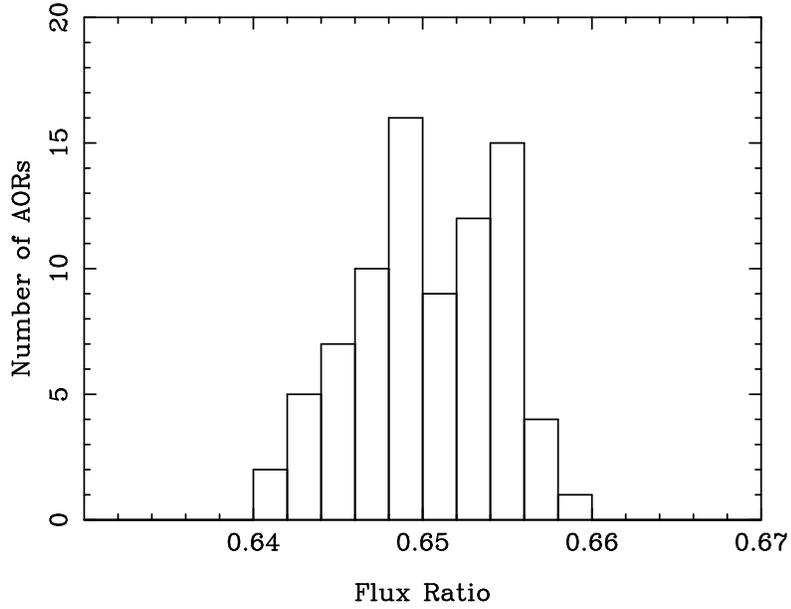}
	\caption{Histogram of ratio of measured flux between channel 2 and channel 1 for the IRAC calibration star, BD +60 1753 over a period of 3 years. While the photometric scatter in any one AOR on this object predicts an uncertainty in this ratio of about 0.002, the measured rms scatter is about 0.004. This systematic difference is primarily due to long term drifts in the instrumental response. See text for details.\label{calibhist}}
    \end{center}
\end{figure}

The systematic uncertainty is measure of the variability in the performance of the instrument, and the spread in values we would obtain by performing the same experiment multiple times. The Observers' Manual quotes a stability for IRAC of $<$1\%. 
To more accurately measure this value, we examined data on \object{BD +60 1753}, a calibration star that has been repeatedly observed by IRAC over the lifetime of the instrument \citep{Reach05a}. We reduced 81 AORs on this star spanning a period of approximately 3.6 years with the same algorithm as we used for our science targets. For each AOR we measured the average of 5 pointings as before, and divided the flux at 4.5\um\ by that at 3.6\um. We show a histogram of this ratio in Figure~\ref{calibhist}. The typical photometric error bar on these measurements is $\delta r$=0.0023, and the rms deviation is 0.0042. As our error terms add in quadrature, the systematic repeatability limit is 0.0035, or 0.54\%.

We add our independent estimate of the systematic uncertainty in quadrature to the photometric uncertainty to get a total (1\,$\sigma$) uncertainty in the ratio, $\delta r$, which is given in the third column of Table~\ref{fluxratio} and shown in Figure~\ref{xsflux}. The {\it excess flux}, $E$ is the difference between $r$ for GD66, 0.6482(62) and the mean value of $r$ for the other two stars, 0.6365(31), or $E$=0.0117(69). The excess flux is only 1.7 times greater than the uncertainty and therefore not statistically significant.

\section{Limits on Planet Mass}
While we did not detect an excess, this does not rule out the evidence from timing measurements, and we can use our observations to constrain the mass of any companion. We can state with 3$\sigma$ confidence that the flux ratio is less than 0.6556. To compare this number to planet models and place an upper bound on the mass of the companion we first need the age of the system. 

Based on the measured surface temperature and gravity \citep[11,989\,K and \logg=8.05][]{Bergeron04}, and comparing to models from \citet{Wood92}, we determine a mass of 0.64(03)\msolar\ and a cooling age of 500\,Myr. The age of the progenitor main-sequence star is more difficult to estimate. Estimates of the initial-final mass relationship (IFMR) for WDs are based on comparing the masses of WDs in open clusters to the main-sequence turn-off age of the cluster itself. Until recently, this relationship was only empirically constrained for young, and hence high mass progenitors. However, newer observations have begun to extend the IFMR down to the relatively lower mass of GD66. Using the linear fits to the IFMR from \citet{Dobbie06}, \citet{Kalirai07} and \citet{Catalan08} we estimate an initial mass of 2.20(46), 2.26(46) and 2.64(57)\msolar\ respectively, which agree within the uncertainties. A theoretical relationship from \citet{Meng07} gives an intermediate value of 2.49(41)\msolar, while the semi-empirical method of \citet{Salaris08} suggests 2.31(22)\msolar. Calculations from \citet{Pols98} indicate that the lifetime of a 2.5\msolar\ star is 830\,Myr. The lifetime of a main sequence star scales roughly as $M^{-2.5}$ in this mass regime, so the calculated main-sequence lifetime of GD66 is 700\,Myr to 1.2\,Gyr. Including the WD cooling age, the total age of the system is 1.2--1.7\,Gyr. 

\begin{figure}
     \begin{center}
 	\includegraphics[angle=270, scale=.45]{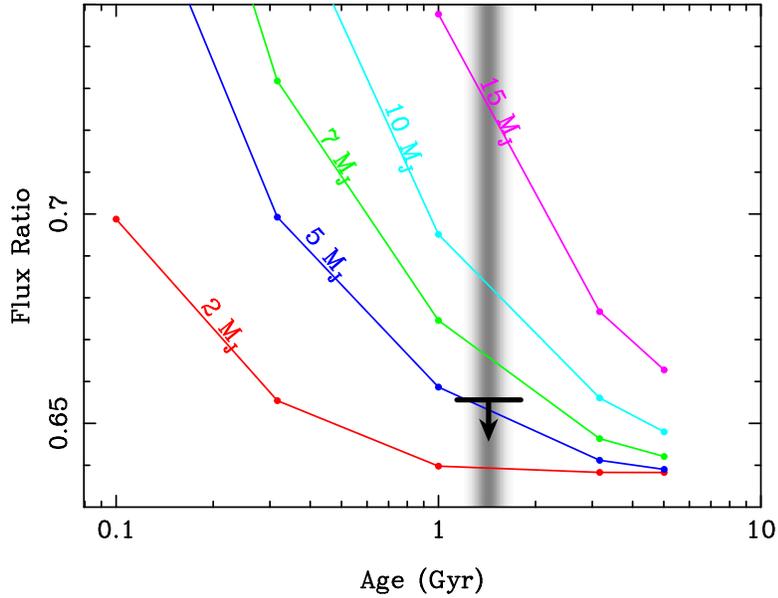}
 	\caption{Mass limits on a planetary companion as a function of the assumed age. The solid lines show the expected flux ratio for WD+planet systems of different ages. The shaded region indicates the best estimate of the age of GD66. The downward arrow indicates the 3$\sigma$ upper bound on the flux ratio.  \label{masslimit}}
     \end{center}
 \end{figure}

 We take atmosphere models of a 12,000\,K DA 
courtesy of D. Koester \citep{Finley97}, and combine them with models of planetary mass objects from \citet{Burrows03}. We perform synthetic photometry on the combined model to determine the flux ratio between channels 1 and 2 for planets of different masses. We repeat the procedure for models with ages from  0.3 to 5\,Gyr and show the results in Figure~\ref{masslimit}. Given our limit on the flux excess, and the uncertainty in the age, the 3$\sigma$ upper bound on the mass of the planet is approximately 5--6\mj.

\section{Additional Time Series Observations}
\begin{figure}
     \begin{center}
 	\includegraphics[angle=270, scale=.5]{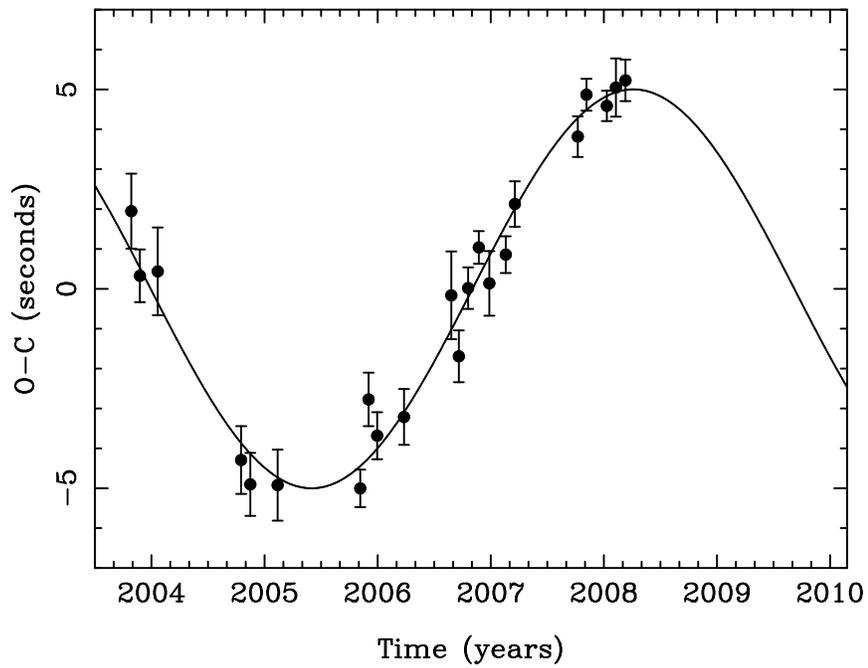}
 	\caption{O-C diagram for GD66. Filled circles are the observed arrival times  of the pulsations. Values greater than zero indicate arrival times that were later than expected based on the assumption of a constant pulsation period and no companions. The solid line is the best fit sinusoid to the data. \label{omc}}
     \end{center}
 \end{figure}

We also present additional ground-based optical time-series observations of GD66 taken with Argos \citep{Nather04} on the 2.1m telescope at McDonald Observatory. A full description of these observations is given in M08, but we repeat a brief summary here. We performed time series differential photometry on GD66 with a continual sequence of exposures over a period of 4--8 hours per night. We divided the lightcurve of the target star by the sum of one of more reference stars to remove the effects of seeing and transparency variations and we corrected our times for the motion of the earth around the sun using the method of \citet{Stumpff80}. We combined all observations over the span of a week into a single data set, and measured the difference between the observed time of arrival of the 302s pulsation mode and that calculated based on the assumption of a constant period (O-C). We show our updated O-C diagram in Figure~\ref{omc}. 

M08 predicted that the orbit would turn over in late 2007 and the slope of the O-C diagram would be negative in 2008. This has clearly not happened and the orbital period is longer than previously expected. Our current best fit parameters are a period of 5.69(30)\,yr, an orbital separation of 2.75(11)\au\ and $m_p.\sin{i}$ = 2.36(17)\mj, where $m_p$ is the mass of the planet and $i$ the inclination of the orbit to the line of sight. With previous experience in mind we are cautious in predicting the future behavior of the orbit, but we note that our current circular best fit orbit appears to be on the verge of turning over.

\section{Discussion}
We did not detect a planet around GD66, but our observations showcase the advantages of targeting WDs in direct detection planet searches. Stars at the end point of their evolution are many magnitudes fainter than when on the main sequence, which more than compensates for the declining flux from the cooling planet. This improved constrast, together with the typically high proper motions of WDs make them ideal candidates for surveys with ground based high contrast imagers being built for large telescopes \citep[e.g. HiCIAO,][]{Tamura06}.

\subsection{Excesses at Long Wavelengths}
We observe slight excesses in channel 4 for L19-2 and ZZ~Ceti. While it is likely that these data are randomly scattered high, it is interesting to speculate on other possible causes. L19-2 and ZZ~Ceti could be explained by a debris disk similar to those found around some metal rich WDs \citep[][and references therein]{Kilic06daz, Farihi08dust}. If we adopt the physically flat, optically thick model given by Eqn. 1 of \citet{Jura03}, and a white dwarf radius, $R_{\mathrm{wd}} = 8.68 \times 10^6$\,m,  we find a best fit temperature for the inner edge of the disk of approximately 220\,K and an orbital separation of $10^{9}$\,m, or 120 $R_{\mathrm{wd}}$. This is significantly more distant than the typical 8-30\,$R_{\mathrm{wd}}$ found by \citet{Jura08irs}, but is consistent with the Roche limit for tidally disrupting an asteroid around a WD \citep{vonHippel07dazd}. However, if such disks existed, we would expect to observe absorption lines of calcium and magnesium in the stellar atmosphere (as we do with all other WDs with debris disks) but this is not the case for these two objects \citep[][S. E. Thompson, priv. comm.]{Koester05}.

\section{Conclusion}
We report on {\it Spitzer} IRAC observations of the white dwarf GD66 in an attempt to directly detect a companion planet. We fail to detect an excess with any statistical significance. However, combined with our ground based timing observations we can now constrain the planet candidate's mass to between 2.4 and $\approx$5--6\mj.

\acknowledgements
We thank Mike Montgomery for donating some of the telescope time used to make ground based observations in this work. We thank Detlev Koester for the use of his atmosphere models. We also thank D. E. Winget, Ted von~Hippel, Mukremin Kilic and Marc Kuchner for their assistance at various stages of this project. This work is based in part on observations made with the Spitzer Space Telescope, which is operated by the Jet Propulsion Laboratory, California Institute of Technology under a contract with NASA. Support for this work was provided by NASA through an award issued by JPL/Caltech.

{\it Facilities:} \facility{MCD:Struve()}; \facility{Spitzer}

\bibliographystyle{apj}
\bibliography{glimmer,mull}

\begin{thebibliography}{}
\expandafter\ifx\csname natexlab\endcsname\relax\def\natexlab#1{#1}\fi
\setlength{\itemsep}{0ex}
\setlength{\itemsep}{-1ex}
\small
\bibitem[{Becklin \& Zuckerman(1988)}]{Becklin88}Becklin, E.~E., \&  Zuckerman, B. 1988, \nat, 336, 656

\bibitem[{Bergeron et~al.(2004)}]{Bergeron04}Bergeron, P., Fontaine, G., Bill\`eres, M.,   	Boudreault, S., \&  Green, E.~M. 2004, \apj, 600, 404

\bibitem[{Burleigh et~al.(2002)}]{Burleigh02}Burleigh, M.~R., Clarke, F.~J., \&  Hodgkin, S.~T. 2002, \mnras, 331, L41

\bibitem[{Burrows et~al.(2003)}]{Burrows03}Burrows, A., Sudarsky, D., \&  Lunine, J.~I. 2003, \apj, 596, 587

\bibitem[{Catal\'an et~al.(2008)}]{Catalan08}Catal\'an, S., Isern, J., Garc\'ia-Berro, E.,  	Ribas, I., Allende Prieto, C., \&  Bonanos, A.~Z. 2008, \aap, 477, 213

\bibitem[{Charbonneau et~al.(2005)}]{Charbonneau05}Charbonneau, D., et~al. 2005, \apj, 626, 523

\bibitem[{Debes et~al.(2007)}]{Debes07cc4}Debes, J.~H., Sigurdsson, S., \&  Hansen, B. 2007, \aj, 134, 1662

\bibitem[{Debes et~al.(2005)}]{Debes05cc2}Debes, J.~H., Sigurdsson, S., \&  Woodgate, B.~E. 2005, \aj, 130, 1221

\bibitem[{Deming et~al.(2006)}]{Deming06}Deming, D., Harrington, J., Seager, S., \&  Richardson L.~J.  , 2006, \apj, 644, 560

\bibitem[{Dobbie et~al.(2006)}]{Dobbie06}Dobbie, P.~D., et~al. 2006, \mnras, 369, 383

\bibitem[{Farihi et~al.(2008a)}]{Farihi08planet}Farihi, J., Becklin, E.~E., \&  Zuckerman, B. 2008a, ArXiv arXiv:0804.0237

\bibitem[{Farihi et~al.(2008b)}]{Farihi08dust}Farihi, J., Zuckerman, B., \&  Becklin, E.~E. 2008b, ArXiv arXiv:0710.0907

\bibitem[{Farihi et~al.(2007)}]{Farihi07gd362}Farihi, J., Zuckerman, B., Becklin, E.~E., \&  Jura M. , 2007,  {\it {15th European Workshop on White Dwarfs}}, ed. Napiwotzki R., \&  Burleigh M.~R., 372, 315 

\bibitem[{Fazio et~al.(2004)}]{Fazio04}Fazio, G.~G., et~al. 2004, \apjs, 154, 10

\bibitem[{Finley et~al.(1997)}]{Finley97}Finley, D.~S., Koester, D., \&  Basri, G. 1997, \apj, 488, 375

\bibitem[{Graham et~al.(1990)}]{Graham90dust}Graham, J.~R., Matthews, K., Neugebauer, G., \&   	Soifer, B.~T. 1990, \apj, 357, 216

\bibitem[{Jura(2003)}]{Jura03}Jura, M. 2003, \apjl, 584, L91

\bibitem[{Jura et~al.(2008)}]{Jura08irs}Jura, M., Farihi, J., \&  Zuckerman, B. 2008, ArXiv 0811.1740

\bibitem[{Kalas et~al.(2008)}]{Kalas08}Kalas, P., et~al. 2008, ArXiv 0811.1994

\bibitem[{Kalirai et~al.(2008)}]{Kalirai07}Kalirai, J.~S., Hansen, B.~M.~S., Kelson, D.~D.,  	Reitzel, D.~B., Rich, R.~M., \&  Richer, H.~B. 2008, \apj, 676, 594

\bibitem[{Kilic et~al.(2006)}]{Kilic06daz}Kilic, M., von Hippel, T., Leggett, S.~K., \&  Winget D.~E.  , 2006, \apj, 646, 474

\bibitem[{Koester et~al.(2005)}]{Koester05}Koester, D., Rollenhagen, K., Napiwotzki, R., Voss, B., Christlieb, N., Homeier, D., \&  Reimers D.     , 2005, \aap, 432, 1025

\bibitem[{Lee et~al.(2008)}]{Lee08}Lee, J.~W., Kim, S.-L., Kim, C.-H., Koch, R.~H.,  	Lee, C.-U., Kim, H.-I., \&  Park, J.-H. 2008, ArXiv 0811.3807

\bibitem[{Livio \& Soker(1984)}]{Livio84}Livio, M., \&  Soker, N. 1984, \mnras, 208, 763

\bibitem[{Marois et~al.(2008)}]{Marois08}Marois, C., Macintosh, B., Barman, T., Zuckerman, B.,  	Song, I., Patience, J., Lafreniere, D., \&  Doyon R. , 2008, ArXiv 0811.2606

\bibitem[{Meng et~al.(2007)}]{Meng07}Meng, X., Chen, X., \&  Han, Z. 2007, ArXiv 0710.2397

\bibitem[{Mukadam et~al.(2006)}]{Mukadam06}Mukadam, A.~S., Montgomery, M.~H., Winget, D.~E.,  	Kepler, S.~O., \&  Clemens, J.~C. 2006, \apj, 640, 956

\bibitem[{Mullally et~al.(2008)}]{Mullally08}Mullally, F., Winget, D.~E., Degennaro, S., Jeffery, E.,  	Thompson, S.~E., Chandler, D., \&  Kepler, S.~O. 2008, \apj, 676, 573

\bibitem[{Mullally(2007)}]{Mullally07thesis}Mullally, F.~R. 2007,  PhD Thesis The University of Texas at Austin

\bibitem[{Nather \& Mukadam(2004)}]{Nather04}Nather, R.~E., \&  Mukadam, A.~S. 2004, \apj, 605, 846

\bibitem[{Pols et~al.(1998)}]{Pols98}Pols, O.~R., Schroder, K.~P., Hurley, J.~R.,   	Tout, C.~A., \&  Eggleton, P.~P. 1998, \mnras, 298, 525

\bibitem[{Probst(1983)}]{Probst83}Probst, R.~G. 1983, \apjs, 53, 335

\bibitem[{Reach et~al.(2005a)}]{Reach05a}Reach, W.~T., et~al. 2005a, \pasp, 117, 978

\bibitem[{Reach et~al.(2005b)}]{Reach05b}Reach, W.~T., Kuchner, M.~J., von Hippel, T., Burrows, A., Mullally, F., Kilic, M., \&  Winget D.~E., 2005b, \apjl, 635, L161

\bibitem[{Reach et~al.(2008)}]{Reach08}Reach, W.~T., Lisse, C., von Hippel, T., \&  Mullally F. , 2008, ArXiv 0810.3276

\bibitem[{Sackmann et~al.(1993)}]{Sackmann93}Sackmann, I.~J., Boothroyd, A.~I., \&  Kraemer K.~E.  , 1993, \apj, 418, 457

\bibitem[{Salaris et~al.(2008)}]{Salaris08}Salaris, M., Serenelli, A., Weiss, A., \&  Miller Bertolami M. , 2008, ArXiv 0807.3567

\bibitem[{Schr\"oder \& Connon~Smith(2008)}]{Schroder08}Schr\"oder, K.~P., \&  Connon~Smith, R. 2008, \mnras, 386, 155

\bibitem[{Silvotti et~al.(2007)}]{Silvotti07}Silvotti, R., et~al. 2007, \nat, 449, 189

\bibitem[{Stumpff(1980)}]{Stumpff80}Stumpff, P. 1980, \aaps, 41, 1

\bibitem[{Tamura et~al.(2006)}]{Tamura06}Tamura, M., et~al. 2006,  {\it {Society of Photo-Optical Instrumentation Engineers (SPIE) Conference Series}}, ed. Kalas, 6269, 34 

\bibitem[{von~Hippel et~al.(2007)}]{vonHippel07dazd}von~Hippel, T., Kuchner, M.~J., Kilic, M., Mullally, F., \&   	Reach, W.~T. 2007, \apj, 662, 544

\bibitem[{Werner et~al.(2004)}]{Werner04}Werner, M.~W., et~al. 2004, \apjs, 154, 1

\bibitem[{Wood(1992)}]{Wood92}Wood, M.~A. 1992, \apj, 386, 539

\bibitem[{Zuckerman et~al.(2007)}]{Zuckerman07}Zuckerman, B., Koester, D., Melis, C., Hansen, B.~M., \&   	Jura, M. 2007, \apj, 671, 872

\end{thebibliography}

\onecolumn
\begin{deluxetable}{lrrllll}
\tablewidth{0pt}
\tablecaption{Observed fluxes \label{allfluxes}}

\tablehead{
    \colhead{Star}&
    \colhead{$J$}&
    \colhead{Integration}&
    \colhead{Ch 1}&
    \colhead{Ch 2}&
    \colhead{Ch 3}&
    \colhead{Ch 4}\\
    \colhead{ }&
    \colhead{(mag)}&
    \colhead{(s)}&
    \colhead{(mJy)}&
    \colhead{(mJy)}&
    \colhead{(mJy)}&
    \colhead{(mJy)}

}

\startdata

GD66 & 15.7 & 6400 & 0.1349(68) & 0.0875(44) & 0.0593(50) & 0.0277(35) \\
L19-2 & 13.6 & 768 & 0.941(47) & 0.600(30) & 0.383(22) & 0.219(12) \\
ZZCeti & 14.4 & 1920 & 0.470(24) & 0.299(15) & 0.189(12) & 0.1105(67) \\
G238-53 & 15.7 & 6400 & 0.1385(69) & 0.0911(46) & 0.0573(39) & 0.0405(26) \\
\enddata
\tablecomments{Integration times are for channels 2 and 4, and are four times shorter for channels 1 and 3. The flux from G238$-$53 is probably overestimated due to the presence of a diffraction spike from a nearby star. The uncertainty in the two least significant digits is given in parentheses. The quoted uncertainties in these measurements include a 5\% absolute calibration uncertainty added in quadrature.}
\end{deluxetable}

\begin{deluxetable}{lcccc}
\tablewidth{0pt}
\tablecaption{Observed fluxes ratios and uncertainties \label{fluxratio}}

\tablehead{
    \colhead{Star}&
    \colhead{Ratio}&
    \colhead{Photometric}&
    \colhead{Systematic}&
    \colhead{Total}\\
    \colhead{ }&
    \colhead{({\footnotesize Ch\,2/Ch\,1}) }&
    \colhead{Uncertainty}&
    \colhead{Uncertainty}&
    \colhead{Uncertainty}
}

\startdata

GD66 & 0.6482 & 0.0051 & 0.0035 & 0.0062\\
ZZCeti & 0.6364 & 0.0026 & 0.0034 & 0.0043\\
L19-2 & 0.6376 & 0.0028 & 0.0034 & 0.0044\\
G238-53 & 0.6574 & 0.0030 & 0.0036 & 0.0046\\

\enddata
\end{deluxetable}

\end{document}